\documentclass[%
 reprint,
 amsmath,amssymb,
 aps,
]{revtex4-2} 

\usepackage{graphicx}
\usepackage{dcolumn}
\usepackage{bm}
\usepackage{bbm}

\begin{document}


\title{Optical interferometer using two-mode squeezed light for enhanced chip-integrated quantum metrology}

\author{Patrick Tritschler}
 \email{patrick.tritschler@bosch-sensortec.com}
\altaffiliation{Bosch Sensortec GmbH, Gerhard-Kindler Stra{\ss}e 9, Reutlingen, 72770, Germany}
\altaffiliation{Institute for Micro Integration (IFM), University of Stuttgart, Allmandring 9b, Stuttgart, 70569, Germany}
\author{Torsten Ohms}%
\altaffiliation{Bosch Sensortec GmbH, Gerhard-Kindler Stra{\ss}e 9, Reutlingen, 72770, Germany}
\author{Andr\'{e} Zimmermann}%
\altaffiliation{Institute for Micro Integration (IFM), University of Stuttgart, Allmandring 9b, Stuttgart, 70569, Germany}
\altaffiliation{Hahn-Schickard, Allmandring 9b, Stuttgart, 70569, Germany}
\author{Fabian Zschocke}%
\altaffiliation{Robert Bosch GmbH, Robert-Bosch-Campus 1, Renningen, 71272, Germany}
\author{Thomas Strohm}
\altaffiliation{Robert Bosch GmbH, Robert-Bosch-Campus 1, Renningen, 71272, Germany}
\author{Peter Degenfeld-Schonburg}%
\altaffiliation{Robert Bosch GmbH, Robert-Bosch-Campus 1, Renningen, 71272, Germany}




\date{\today}

\begin{abstract}
	This work discusses the possibility of using two-mode squeezed light to improve the performance of existing sensor technology with the focus on its miniaturization under realistic losses. Therefore, we analyze a system consisting of a part for the two-mode squeezed light generation, a sensor region and a detection stage. Based on a general four-wave mixing (FWM) Hamiltonian caused by the third order susceptibility, we formulate linearized equations that describe the FWM process below the threshold and are used to analyze the squeezing quality of the generated optical signal and idler modes. For a possible realization, the focus is set on the chip-integrated generation using micro-ring resonators. To do so, the impact of the design and the pump light are considered in the derived equations. These equations are used to analyze the usage of two-mode squeezed light in quantum metrology and the application in a Mach-Zehnder interferometer (MZI).
	 Due to the impact of losses in realistic use cases, we show that the main usage is for small and compact devices, which can lead to a quantum improvement up to a factor of ten in comparison of using coherent light only. This enables the use of small squeezing-enhanced sensors with a performance comparable to larger classical sensors.
\end{abstract}

\maketitle


\section{\label{sec:introduction} Introduction}

Quantum metrology has the potential to use completely new sensing principles, to enable new use cases and to increase the sensitivity of existing measurement technology \cite{Nawrocki2019}. The main motivation for quantum metrology in comparison to its classical counterpart is that quantum-sensors can theoretically overcome the shot noise limited with $1/\sqrt{N}$ and achieve the Heisenberg limit with $1/N$ where $N$ denotes the number of probes which can be for example the number of photons or the number of measurement repetitions \cite{PhysRevLett.96.010401, Giovannetti2011}. However, in reality this is very challenging to achieve. The reason is that quantum states like single-photons are heavily affected by losses and decoherence effects which makes it difficult to build a Heisenberg-limited quantum sensor \cite{PhysRevLett.102.040403, DEMKOWICZDOBRZANSKI2015345}. Despite this, there are applications which make use of a quantum enhancement, like the famous example of gravitational-wave detection. Thereby, single-mode squeezed-light is used to increase the phase-sensitivity of a huge interferometer to detect extremely small signals caused by gravitational waves \cite{PhysRevLett.116.061102, Vahlbruch_2010}. \\ %
In this work, we show the potential of using two-mode squeezed light generated by FWM for enhanced optical interferometry. The FWM process is utilized in many applications like the creation of frequency-combs for sensing and communication \cite{Chang2020, Fortier2019} or squeezed state generation for optical quantum computing \cite{Arrazola2021}. We want to use it to generate two-mode squeezed light for sensing applications. The main focus is on a realistic quantum metrology platform that is realizable with existing technology in photonic integrated circuits. \\
We see big potential of quantum metrology for the miniaturization of existing sensor concepts and to realize them in chip-integrated applications. Therefore, we show theoretical derivations of the realistic sensor performance, which includes the equations required to perform a proper design with typical losses and manufacturing technologies. \\
The paper is structured as follows. First, the system of interest is discussed in the next section \ref{cha:system_overview}, which consists of the state preparation for generating a signal and an idler mode via FWM, an MZI as a sensor region and a detection stage. Afterwards, each part of the system is discussed in detail. In chapter \ref{cha:four-wave-mixing}, we start with the discussion of the FWM process in a ring-resonator and proceed with the interaction of the environment in chapter \ref{cha:dissipation} to determine the output modes in chapter \ref{cha:input-output} using input-output theory. Subsequently, section \ref{cha:mach-zehnder} discusses how the generated modes are used to detect a phase shift in a lossy MZI. The main results are shown in section \ref{cha:sensingLimit} in which it is discussed that the combination of squeezed and coherent light can lead to an quantum-enhancement and that the sensor performance can be improved by an order of magnitude. However, due to optical losses, squeezed light is only useful for a compact or low-loss sensor concept, which leads to an advantage as a two-mode squeezed chip-integrated optical MZI.

\begin{figure*}
	\includegraphics[scale=0.37]{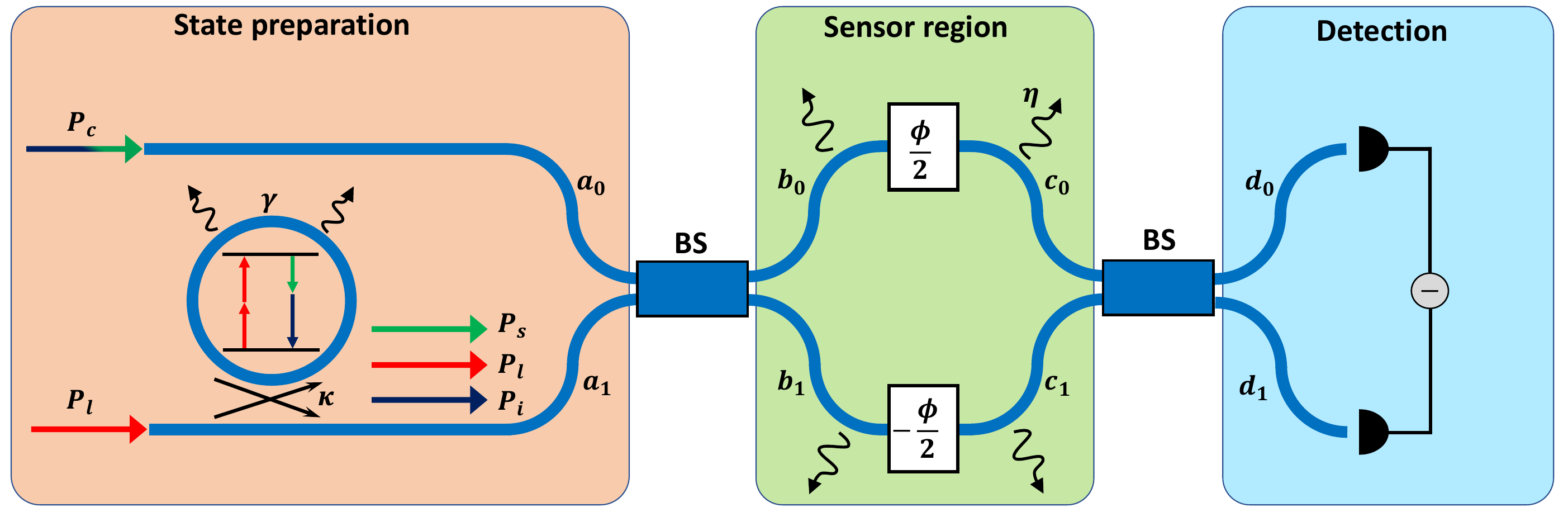}
	\caption{\label{fig:MZI}Schematic of a MZI which consists of a state preparation (on the left; orange), sensor region (in the middle; green) and a detection stage (on the right; blue).}
\end{figure*}

\section{System overview}
\label{cha:system_overview}
The system of interest is shown in Figure \ref{fig:MZI} and corresponds to an MZI that is pumped by two-mode squeezed and coherent light. It consists overall of three stages with the state preparation, the sensor region and the detection stage. In the state preparation stage, the system has two inputs in which two coherent light sources can be inserted which are denoted by their power $P_l$ and $P_c$. While $P_c$ is sent directly to the MZI and corresponds to the input mode $a_0$, $P_l$ is used to pump a resonator to generate squeezed signal and idle pairs with powers $P_s$ and $P_i$ corresponding to the other input mode of the MZI $a_1$. Thereby, $P_l$ consists of light with the frequency $\omega_l$ and the generated signal and idler light of $\omega_s$ and $\omega_i$ with $\omega_s+\omega_i=2\omega_l$. \\
The resonator is a ring-shaped waveguide that is connected to the straight waveguide and forms a cavity. The light can couple from the straight waveguide into the ring and the other way around by evanescent coupling which can be described by the coupling rate $\kappa$. Losses appear inside of the resonator which are described by the loss rate $\gamma$. The physical relations to the parameters $\kappa$ and $\gamma$ as a function of the design variations of the ring are explained in more detail in Appendix \ref{cha:physRel}.  Both input modes $a_0$ and $a_1$ are mixed at a multi-mode interferometer (MMI) which acts as a beam splitter (BS) and are send into the sensor region of the MZI. For a useful interaction between $P_s$, $P_i$ and $P_c$, it is required that $P_c$ consists of the same frequency components as $P_s$ and $P_i$ with $\omega_s$ and $\omega_i$. This is denoted by the colors (gray scales) of each in the figure \ref{fig:MZI}. Thus, the transmitted power $P_l$ is not required anymore and is filtered out in the following. In this second stage, the light is affected by losses $\eta$ as well as a phase shift $\Phi$ which should be detected and is assumed to be equally distributed between the two light paths. 
At the end of the sensor region, $c_0$ and $c_1$ are mixed at a BS and finally detected at the detection stage using intensity difference detection. \\
In the following, we first describe the FWM process inside of the resonator and determine the expectation values of the signal and idler modes that form the mode $a_1$. Afterwards, we use these results to determine the minimum detectable phase change of the system.

\section{Four-Wave Mixing}
\label{cha:four-wave-mixing}
FWM is a well-known non-linear optical process which describes the effect of two pump modes $a_p(t)$ being absorbed by a material with a significant third-order susceptibility coefficient $\chi^{(3)}$ while emitting two new modes which are called signal $a_s(t)$ and idler $a_i(t)$ with energy conservation $\omega_i+\omega_s=2\omega_p$. For simplicity we write $a_i$ for the time dependent operator. Following \cite{PhysRevA.92.033840}, the Hamiltonian describing FWM inside a cavity is given in the rotating-wave approximation by
\begin{equation}\label{equ:fwm_h}
\begin{aligned}
	H_{\mathrm{FWM}}  = & \hbar\omega_p a_p^{\dagger}a_p  + \hbar\omega_s a_s^{\dagger}a_s + \hbar \omega_i a_i^{\dagger}a_i & \\
	& +i\hbar g\left(a_p a_p a_s^{\dagger} a_i^{\dagger} - a_p^{\dagger} a_p^{\dagger} a_s a_i\right) &
\end{aligned}
\end{equation}
with the nonlinear gain $g$, which is directly proportional to 
$\chi^{(3)}$ and is discussed further in appendix \ref{cha:fwm_gain}. The first line of equation \ref{equ:fwm_h} corresponds to the relevant resonance frequencies in the resonator $\omega_p$, $\omega_s$, $\omega_i$. The desired FWM process is represented in the second line and denotes the non-linear interaction in which two pump modes are either absorbed or generated in exchange with one signal and idler mode. In practice, effects such as self-phase modulation and cross-phase modulation also occur, where a mode influences its own and the resonant frequencies of other modes. However, both are neglected in this paper.\\
A linearization of the FWM part in the Hamiltonian is then performed, assuming the pump mode inside of the resonator to be in a coherent state with the complex amplitude $\langle a_p \rangle = \alpha_{p}$, which leads to the following Hamiltonian
\begin{equation}\label{equ:fwm_h_lin}
\begin{aligned}
	H_{\mathrm{FWM},\mathrm{lin}} = &\hbar(\omega_p \alpha_{p}^{*}\alpha_{p} + \omega_s a_s^{\dagger}a_s + \omega_i a_i^{\dagger}a_i)& \\
	& +  \frac{i \hbar}{2} ( \sigma a_s^{\dagger} a_i^{\dagger} - \sigma^* a_s a_i)& \\
\end{aligned}
\end{equation}
where $\sigma=2 g \cdot \alpha_{p}^2$ is the injection parameter. Note that the conservation of energy must be fulfilled in the FWM process, but this is not always the case due to dispersion. To include this effect similar as in the in \cite{PhysRevA.92.033840}, the injection parameter can include a FWM detuning $\Delta_{\sigma}=\omega_s+\omega_i-2\omega_p$ and a certain phase with 
\begin{equation}
	\sigma =  |\sigma| \cdot e^{-i(\Delta_{\sigma}t - \phi_{\sigma})}.
\end{equation} 
This detuning should be zero for the best performance and can this be achieved by dispersion engineering of the waveguide geometry \cite{FujiiTanabe+2020+1087+1104}.

\section{Dissipation}
\label{cha:dissipation}
\begin{figure*}
	\includegraphics[scale=0.36]{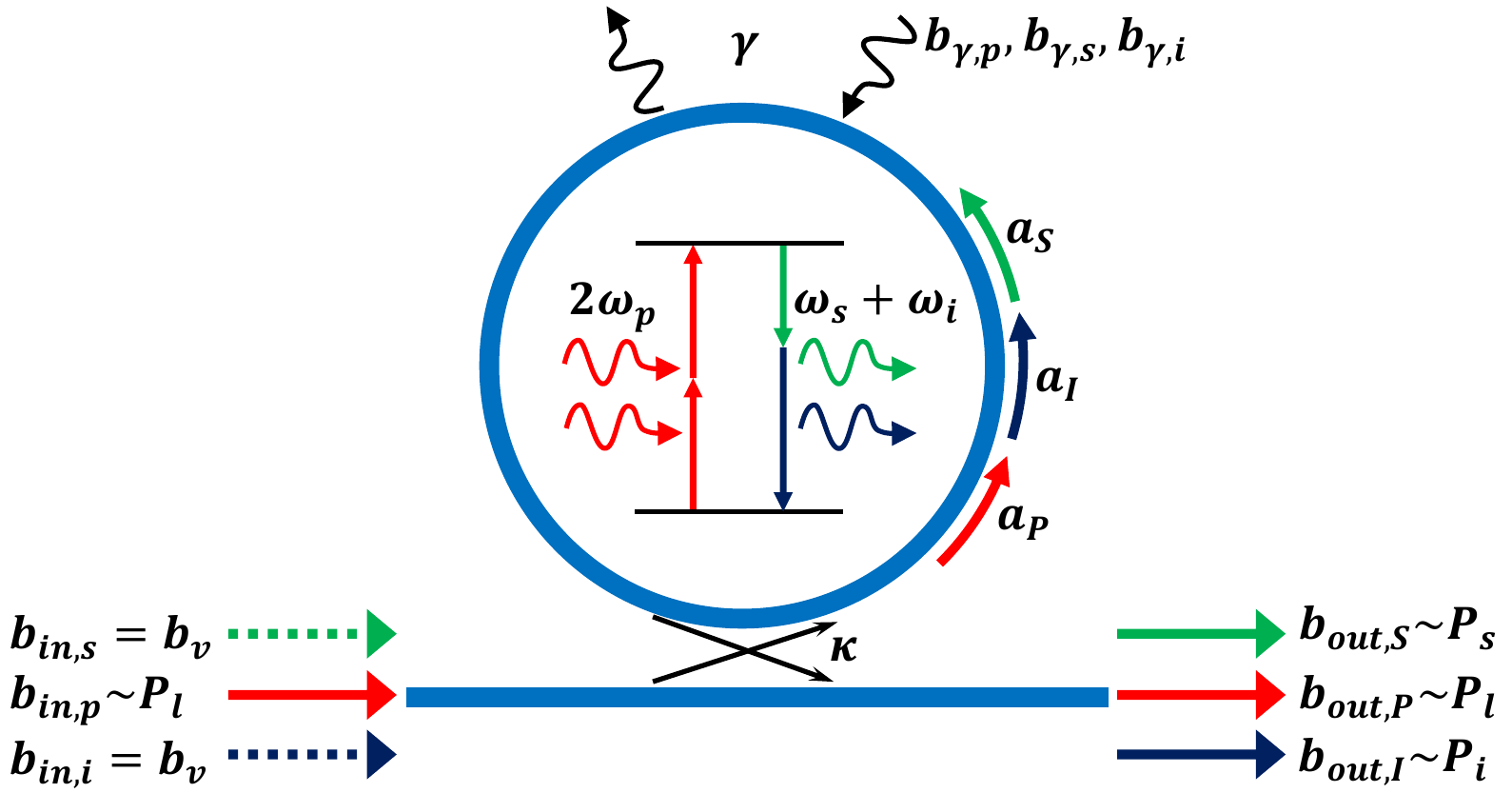}
	\includegraphics[scale=0.66]{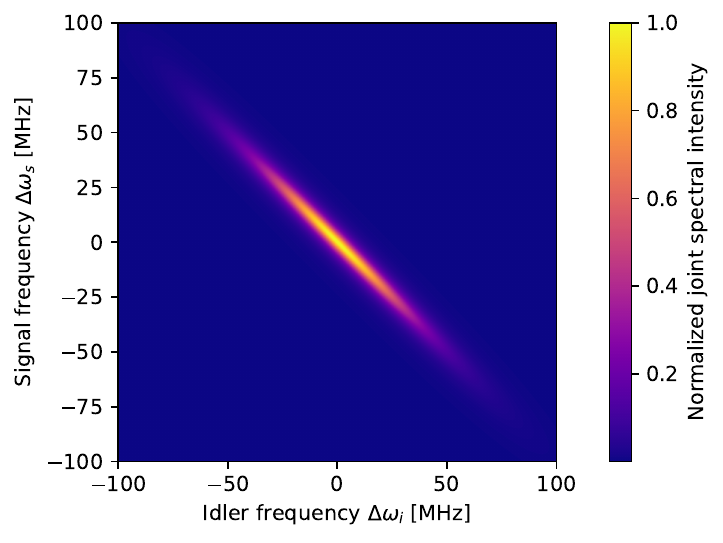}
	\caption{\label{fig:ring_squeezing}\textbf{Left:} Ring resonator which is used to generate squeezed light via FWM. A pump mode with the power $P_p$ is coupled via the parameter $\kappa$ into the cavity, in which a strong pump field generates a signal and idler mode, which couple out of the cavity. Occurring losses inside of the resonator are considered with $\gamma$. \textbf{Right:} Joint spectral intensity between the signal and idler mode calculated using equation \ref{equ:jsi} with $X=0.01$, $n_{\mathrm{eff}}=1.801$, $n_g=2.10087$, $\alpha_{loss}=0.23$ $1/\mathrm{m}$, $L=2\pi\cdot220\cdot10^{-6}$ m, $\omega_p=1550\cdot10^{-6}$ m, $n_2=2.4\cdot10^{-19}$ $\mathrm{m}^2/\mathrm{W}$, $A_{\mathrm{eff}}=1.05564\cdot10^{-12}$ $\mathrm{m}^2$, $\sigma=0.995\cdot\sigma_{\mathrm{th}}$ and normalized by the maximum value of $2.98\cdot10^{9}$.}
\end{figure*}
To analyze the behaviour of FWM inside of a realistic system, it is important to determine the equations of motion of the modes inside of the resonator under the influence of losses and a drive. Therefore, we focus on the desired resonator structure that is shown in Figure \ref{fig:ring_squeezing}. A laser pump mode with the amplitude $\alpha_l=|\alpha_l|\cdot \exp(i\phi_{l})=\sqrt{P_l/\hbar\omega_l}\cdot \exp(i\phi_{l})$ and the laser power $P_l$ and phase $\phi_l$ is coupled from the straight waveguide to the resonator over $\kappa$. An enhanced cavity pump field $\alpha_{p}$ is achieved if the resonance condition is fulfilled with $\omega_r = 2\pi m c / n_{\mathrm{eff}}L$ and the mode index $m=1, 2, ...$, the effective index $n_{\mathrm{eff}}$ and the resonator length $L$ \cite{MicroRingTheory}. In the following we assume that this is fulfilled with $\omega_l=\omega_p$. Since we focus on a microscopic device, the integrated resonator is the preferred structure. The advantage is that due to the field enhancement and the small mode volume, less laser power $P_l$ is required to achieve FWM and to generate squeezed signal and idler modes. \\
Due to imperfections of the waveguide and mostly caused by scattering, cavity energy is lost which is described by the loss rate $\gamma$. Both rates $\kappa$ and $\gamma$ describe the photons leaving or entering the cavity due to the respective effect. To include the losses in the equations, it is important to describe the interaction of the ring resonator system with the environment which consists of the straight waveguide and the bath around the ring resonator. From the straight waveguide, the input modes $b_{in,s}$, $b_{in,i}$ and $b_{in,p}$ couple from the waveguide in the ring, while the resonator modes couple from the ring into the waveguide. The bath modes $b_{\gamma,s}$, $b_{\gamma,i}$ and $b_{\gamma,p}$ interact through the losses $\gamma$ with the resonator modes. Thereby, we assume the waveguide modes, $b_{in,s}$ and $b_{in,i}$ and the bath modes to be in the vacuum state, while $b_{in, p}$ represents the coherent pump $\alpha_p$. In appendix \ref{cha:expResults}, we also discuss the case of a coherent part in $b_{in,s}$ and $b_{in,i}$.\\
It is well known that the interaction between the system and the environment can be described in dependency of the decay rates with $\sqrt{\kappa} b_{in,\mu}$ and $\sqrt{\gamma} b_{\gamma, \mu}$ and the modes $\mu=p,s,i$ \cite{Collet84, Gardiner85, scully_zubairy_1997}. This leads to the following interaction Hamiltonian in the rotating-wave approximation.
\begin{equation}\label{equ:interaction_H}
\begin{aligned}
	H_{\mathrm{int}} \approx i \hbar &\biggl(\sqrt{\frac{\kappa}{2\pi}} \int_{-\infty}^{\infty} d\omega \sum_{\mu={i, s, p}} [ a_\mu b_{in,\mu}^\dagger\left(\omega\right) - a_\mu^\dagger b_{in,\mu}\left(\omega\right)] & \\
	& +  \sqrt{\frac{\gamma}{2\pi}} \int_{-\infty}^{\infty} d\omega \sum_{\mu={i, s, p}} [ a_\mu b_{\gamma,\mu}^\dagger\left(\omega\right) - a_\mu^\dagger b_{\gamma,\mu}\left(\omega\right) ] \biggr)&
\end{aligned}
\end{equation}
Note that if the wavelength difference for the signal, idler and pump is significant, it is important to distinguish between the coupling and loss rates of each. Both rates would then require the corresponding indices.\\
With equation \ref{equ:fwm_h_lin} and \ref{equ:interaction_H}, it is possible to derive the quantum Langevin equation of motion in the Heisenberg picture describing the intra-cavity modes in dependency of the bath and waveguide modes which can be derived for our system using \cite{Lambropoulos2007, Gardiner85}
\begin{eqnarray}\label{equ:eqm_cavityModes}
\begin{aligned}
	\frac{d a_{\mu}(t)}{dt} = &- \frac{i}{\hbar} [a_{\mu}(t) , H_{\mathrm{FWM},\mathrm{lin}}]&\\
	 &- \frac{\gamma+\kappa}{2} a_{\mu}(t) + \sqrt{\kappa} b_{in,\mu}(t) + \sqrt{\gamma} b_{\gamma,\mu}(t).&
\end{aligned}
\end{eqnarray}
Thereby, the modes $b_{in,\mu}(t)$ and $b_{\gamma,\mu}(t)$ of equation \ref{equ:eqm_cavityModes} are in units of $\sqrt{\mathrm{Hz}}$ and are connected by a Fourier transformation with the corresponding modes of equation \ref{equ:interaction_H}. This is explicitly denoted by the dependency of either $t$ or $\omega$. First, we solve equation \ref{equ:eqm_cavityModes} for the pump mode. The fluctuations in the pump mode can be neglected in this case due to the strong coherent background. Setting $\langle a_p \rangle = \alpha_{p}$, this leads to the following equation of motion
\begin{equation}
	\frac{d}{d t} \alpha_{p}(t) = -i \omega_p \alpha_{p}(t) - \frac{\gamma + \kappa}{2} \alpha_{p}(t) + \sqrt{\kappa} \alpha_l(t).
\end{equation}
This equation is solved by performing a Fourier-transformation, which leads to
\begin{equation}
	\alpha_{p}\left(\omega\right) = \frac{\sqrt{\kappa} \alpha_l\left(\omega\right)}{(\gamma + \kappa)/2 - i(\omega - \omega_p)}
	\label{equ:cavity_pumpPhotons}
\end{equation}
with the detuning from the resonance wavelength $\Delta_p=\omega - \omega_p$, which is an important part if effects like self-phase modulation or optical bi-stability are considered. Note that while $\alpha_l$ has the units of $\sqrt{\mathrm{Hz}}$, the intracavity mode $\alpha_{p}$ is unitless. The reason is that while the modes outside of the resonator correspond to a photon flux in units of $\sqrt{\mathrm{Hz}}$, the mode $\alpha_{p}$ describes the field of the resonator over the whole mode volume. It is described in appendix \ref{cha:physRel} how to transfer $\alpha_{p}$ back to a physical unit which corresponds to a photon flux. For the intra-cavity signal and idler modes, the fluctuations are relevant and the equation of motion can be given as
\begin{eqnarray}\label{equ:motion_in_1}
\begin{aligned}
\frac{d a_{s/i}(t)}{dt} =& - i \omega_{s/i} a_{s/i}(t) + \frac{\sigma}{2} a_{i/s}^{\dagger}(t) - \frac{\gamma}{2} a_{s/i}(t) - \frac{\kappa}{2} a_{s/i}(t)&  \\
& + \sqrt{\kappa} b_{in, s/i}(t) + \sqrt{\gamma} b_{\gamma, s/i}(t)& \\
\end{aligned}
\end{eqnarray}

\section{Input-Output Theory}
\label{cha:input-output}
To be able to determine the performance of sensor systems using the generated two-mode squeezed states, it is required to 
solve the equations of motions \ref{equ:motion_in_1} and to determine the expectation values of the output modes that depend on the geometrical and pump parameters. In the following, we derive and validate the equations and analyze the squeezing of the output modes.

\subsection{Linearized outra-cavity equations}
To derive the output modes $b_{out}$ of Figure \ref{fig:ring_squeezing}, we use the boundary condition that has already been derived in \cite{Collet84, Gardiner85} and that has been proven to be valid for a vacuum and coherent input modes with
\begin{equation} \label{equ:boundary_condition}
b_{in, s/i}(t) = \sqrt{\kappa} a_{s/i}(t) - b_{out, s/i}(t).
\end{equation}
Thereby, $b_{out, s/i}$ correspond to the signal and idler output modes, which couple from the ring to the straight waveguide and has the same units as $b_{in, s/i}$ with $\sqrt{\mathrm{Hz}}$. This leads to the following equations of motion describing the intra-cavity field using the output operator
\begin{eqnarray} \label{equ:motion_out_1}
\begin{aligned}
\frac{d a_{s/i}(t)}{dt}  = &- i \omega_{s/i} a_{s/i}(t) + \frac{\sigma}{2} a_{i/s}^{\dagger}(t) - \frac{\gamma}{2} a_{s/i}(t) + \frac{\kappa}{2} a_{s/i}(t)&\\
 &- \sqrt{\kappa} b_{out, s/i}(t) + \sqrt{\gamma} b_{\gamma,s/i}(t) &
\end{aligned}
\end{eqnarray}
For a more compact description, the equations \ref{equ:motion_in_1} and \ref{equ:motion_out_1} can be given in matrix notation as follows for the input and output operators.
\begin{eqnarray}
	\frac{d}{dt} \mathbf{A}(t) &=& (\mathbf{K}+\frac{\kappa}{2} \mathbbm{1}) \mathbf{A}(t) +  \sqrt{\kappa} \mathbf{B_{in}}(t) + \sqrt{\gamma} \mathbf{B_{\gamma}}(t), \label{equ:eqm_t_in}\\ 
	\frac{d}{dt} \mathbf{A}(t) &=& (\mathbf{K}-\frac{\kappa}{2} \mathbbm{1}) \mathbf{A}(t) - \sqrt{\kappa} \mathbf{B_{out}}(t) + \sqrt{\gamma} \mathbf{B_{\gamma}}(t).\label{equ:eqm_t_out}
\end{eqnarray}
Thereby, $\mathbbm{1}$ is the $4\times4$ identity matrix and we use the following notations for the other matrices.
\begin{eqnarray}
	\mathbf{B_{in}} = \left(\begin{array}{c}
	b_{in,s}\\
	b_{in,s}^{\dagger} \\
	b_{in,i} \\
	b_{in,i}^{\dagger}
	\end{array}\right), \quad
	\mathbf{B_{\gamma}} = \left(\begin{array}{c}
	b_{\gamma, s,}\\
	b_{\gamma,s}^{\dagger} \\
	b_{\gamma,i} \\
	b_{\gamma,i}^{\dagger}
	\end{array}\right),
\end{eqnarray}
\begin{eqnarray}
\mathbf{A} = \left(\begin{array}{c}
a_s\\
a_s^{\dagger} \\
a_i \\
a_i^{\dagger}
\end{array}\right), \quad
\mathbf{B_{out}} = \left(\begin{array}{c}
b_{out,s}\\
b_{out,s}^{\dagger} \\
b_{out,i} \\
b_{out,i}^{\dagger}
\end{array}\right),	
\end{eqnarray} 
\begin{eqnarray}
	\mathbf{K} = \left(\begin{matrix} 
	-i\omega_s - \frac{\gamma}{2} & 0 & 0 & \frac{\sigma}{2} \\
	0 & i \omega_s - \frac{\gamma}{2} & \frac{\sigma^*}{2} & 0 \\
	0 & \frac{\sigma}{2} &  - i \omega_i - \frac{\gamma}{2} & 0 \\
	\frac{\sigma^*}{2} & 0 & 0 & i \omega_i - \frac{\gamma}{2} \\
	\end{matrix} \right).
\end{eqnarray}
Performing a Fourier-transformation on equation \ref{equ:eqm_t_in} and \ref{equ:eqm_t_out}, leads to the following matrix notations of the equation of motion in the frequency domain
\begin{eqnarray}
	[\mathbf{\Omega}-\mathbf{K}+\frac{\kappa}{2} \mathbf{I_4}] \mathbf{A(\omega)} &=& \sqrt{\kappa} \mathbf{B_{in}}(\omega) + \sqrt{\gamma} \mathbf{B_{\gamma}}(\omega),\\ 
	\mathbf{} [\mathbf{\Omega}-\mathbf{K}-\frac{\kappa}{2} \mathbf{I_4}] \mathbf{A(\omega)} &=& - \sqrt{\kappa} \mathbf{B_{out}}(\omega) + \sqrt{\gamma} \mathbf{B_{\gamma}}(\omega),
\end{eqnarray}
where $\mathbf{\Omega}$ is the frequency matrix with
\begin{eqnarray}
	\mathbf{\Omega} = \left(\begin{matrix} 
	-i \omega  & 0 & 0 & 0 \\
	0 & i \omega   & 0 & 0 \\
	0 & 0 &  -i \omega  & 0 \\
	0 & 0 & 0 & i \omega  \\
\end{matrix} \right)
\end{eqnarray}
This is now a linear equation system which can be re-arranged to eliminate the intra-cavity modes and to determine the output modes of the system using only the bath modes, waveguide modes and the system operators with
\begin{eqnarray}\label{equ:outputMode}
\begin{aligned}
	\mathbf{B_{out}}(\omega) = &- \frac{1}{\sqrt{\kappa}}  \Bigl\lbrack[\mathbf{\Omega}-\mathbf{K}-\frac{\kappa}{2} \mathbf{I_4}][\mathbf{\Omega}-\mathbf{K}+\frac{\kappa}{2} \mathbf{I_4}]^{-1} &\\
	&\cdot\left( \sqrt{\kappa}\mathbf{B_{in}}(\omega) + \sqrt{\gamma} \mathbf{B_{\gamma}}(\omega) 
	\right)- \sqrt{\gamma} \mathbf{B_{\gamma}}(\omega) \Bigr\rbrack. &
\end{aligned}
\end{eqnarray}
Note that by subtracting $\mathbf{\Omega}$ from $\mathbf{K}$, we introduce the detunings of the signal and idler mode with $\Delta_i = \omega - \omega_i$ and $\Delta_s = \omega - \omega_s$, which are zero for a perfect cavity design and FWM phase-matching. \\
By using equation \ref{equ:outputMode}, it is possible to determine the expectation values of the output modes. To achieve this, we make use of the canonical commutation relations of the vacuum modes $b_v$ which leads to the following expectation values \cite{zoller1997quantum, wiseman_milburn_2009}
\begin{eqnarray} \label{equ:vacuum_exp}
\begin{aligned}
    \langle b_v( t) \rangle &=  \langle b_v^{\dagger}(t) \rangle = \langle b_v( t) b_v( t' ) \rangle = \langle b_v^{\dagger}( t) b_v( t' ) \rangle = 0& \\
	\langle b_v\left( t \right) b_v^{\dagger}( t' ) \rangle &= \delta(t-t').&
\end{aligned}
\end{eqnarray}
While the results for all expectation values are given in appendix \ref{cha:expResults}, we want to give particular emphasis to the expectation value of the output signal photon number, which are 
\begin{equation}\label{equ:photon_n}
\begin{aligned}
	\langle b_{out,s}^\dagger \left(\omega \right) b_{out,s} \left(\omega' \right) \rangle =& \frac{4\sigma^2\kappa\Gamma}{\Xi -2\sigma^2\Gamma^2} \delta(\omega - \omega') \\ \overset{\Delta_i=\Delta_s=0}{=} &\frac{4\sigma^2\kappa\Gamma}{(\Gamma^2 - \sigma^2)^2} \delta(\omega - \omega')
\end{aligned}
\end{equation}
with 
\begin{equation}
\Xi = (4\Delta_i\Delta_s-\sigma^2)^2 + 4\Gamma^2(\Delta_i^2+\Delta_s^2) + \Gamma^4
\end{equation}
and the total losses of the cavity modes $\Gamma=\gamma+\kappa$. It is important to note that equation \ref{equ:photon_n} diverges for $\sigma=\Gamma:=\sigma_{\mathrm{th}}$, which corresponds to the threshold of the FWM process and is discussed in more detail in appendix \ref{cha:physRel}. 

\subsection{Mean-field theory}
\begin{figure*}
	\includegraphics[scale=0.64]{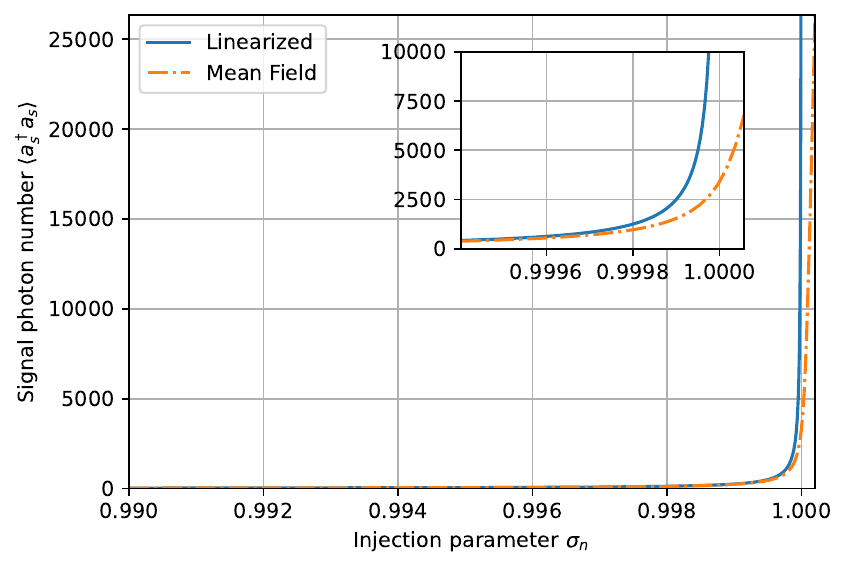}
	\includegraphics[scale=0.64]{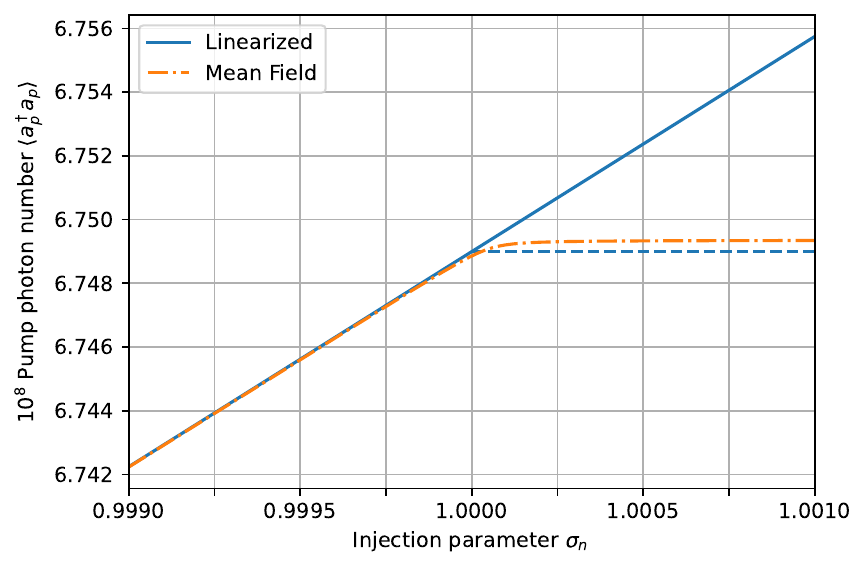}
	\caption{\label{fig:Mean_field_error} \textbf{Left:} Signal photon number $\langle a_p^\dagger a_p \rangle$ for the linearized approach (solid blue) and MF (dashed dotted orange) \textbf{Right:} Pump photon number $\langle a_p^\dagger a_p \rangle$ for the linearized approach (solid blue), the solution of the linearized approach at the threshold (dashed blue) and MF (dashed dotted orange). For both the following values are used: $\Delta_i=\Delta_s=\Delta_p=0$, $X=0.01$, $n_{\mathrm{eff}}=1.801$, $n_g=2.10087$, $\alpha_{loss}=0.23 $ $1/\mathrm{m}$, $L=2\pi\cdot220\cdot10^{-6}$ m, $\lambda_p=1550\cdot10^{-6}$ m, $n_2=2.4\cdot10^{-19} $ $\mathrm{m}^2$/W, $A_{\mathrm{eff}}=1.05564\cdot10^{-12}$ $\mathrm{m}^2$.}
\end{figure*}
The divergence in equation \ref{equ:photon_n} is not physical and thus, it is necessary to know the range of $\sigma$ in which the derived equations are valid as well as to validate the linearized Hamiltonian approach. For this, the results of the expectation values are determined using the mean-field theory (MF). Therefore, we use the following drive term
\begin{equation}
	H_{\mathrm{d}} = i\hbar (\sqrt{\kappa} \alpha_l a_p^\dagger - \sqrt{\kappa}\alpha_l^* a_p)
\end{equation}
as well as the FWM Hamiltonian from equation \ref{equ:fwm_h} and the linearized one in equation  \ref{equ:fwm_h_lin} to derive a master equation describing the intra-cavity modes for the MF as well as the linearized approach with
\begin{align}
	\frac{d \rho_{\mathrm{MF}} }{d t} &= - \frac{i}{\hbar} [H_{\mathrm{FWM}} + H_{\mathrm{d}}, \rho_{\mathrm{MF}}]& \nonumber \\ &+ \sum_{j={i, s, p}} \Gamma [a_j \rho_{\mathrm{MF}} a_j^\dagger - \frac{1}{2} (a_j^\dagger a_j\rho_{\mathrm{MF}} + \rho_{\mathrm{MF}} a_j^\dagger a_j)]& \label{equ:mf_ME}\\
	\frac{d \rho_{\mathrm{lin}} }{d t} &= - \frac{i}{\hbar} [H_{\mathrm{FWM,lin}}, \rho_{\mathrm{lin}}]& \nonumber \\ &+ \sum_{j={i, s}} \Gamma [a_j \rho_{\mathrm{lin}} a_j^\dagger - \frac{1}{2} (a_j^\dagger a_j\rho_{\mathrm{lin}} + \rho_{\mathrm{lin}} a_j^\dagger a_j)]. \label{equ:lin_ME} &
\end{align}
In comparison to the MF master equation, the pump mode does not appear in the linearized equation. Each of the master equations leads to a coupled equation system which can be solved to determine the expectation values using a numerical approach. However, for the MF approach, nonlinear terms arise which are solved by decoupling the expectation values of the pump mode from the signal and idler modes with for example $\langle a_p^\dagger a_p a_s a_i \rangle = \langle a_p^\dagger a_p \rangle \langle a_s a_i \rangle$. Thus, MF is a more realistic view since the pump mode still appears in the Hamiltonian as an operator. Knowing that MF is not perfect as well, it has been shown that it is very accurate below the threshold \cite{Navarrete-Benlloch:14, PhysRevA.91.053850, PhysRevA.93.023819}. Thus, we can use it to validate the range of $\sigma$. \\
Values of a real system are used to perform a simulation of an optical silicon nitride ($\mathrm{Si_3N_4}$) waveguide with the height of $800$ nm, a width of $1.2$ $\mu$m and a bend radius of $220$ $\mathrm{\mu}$m using a mode solver at a pump wavelength of $\lambda_p=1.55$ $\mu$m. This dimension is in the range where squeezing generation with a ring resonator has already been demonstrated \cite{doi:10.1126/sciadv.aba9186}. The nonlinear index is approximated to be $2.4\cdot10^{-19}$ $\mathrm{m}^2/\mathrm{W}$ based on \cite{Ikeda:08} which leads to $g=1.5$ Hz. Losses of $\alpha_{loss}=1 \mathrm{dB}/\mathrm{m}$ and a ring transmission of $0.01$ are assumed. This lead to $\gamma=38.3$ MHz and $\kappa=1208$ MHz.\\
The results for the signal photon number $\langle a_s^\dagger a_s \rangle$ and the pump photon number $\langle a_p^\dagger a_p \rangle$ are shown in Figure \ref{fig:Mean_field_error} as a function of the normalized injection parameter $\sigma_n=\sigma/\sigma_{\mathrm{th}}$ which is exactly one at the threshold. While for MF $\langle a_p^\dagger a_p \rangle$ is determined using the master equation, the linearized approach is solved using equation \ref{equ:cavity_pumpPhotons}. It can be seen that pump depletion appears at $\sigma_c = 1$ for MF while $\langle a_p^\dagger a_p \rangle$ rises for the linearized approach. For $\langle a_s^\dagger a_s \rangle$ it can be seen that the linearized approach diverges at $\sigma_n=1$, while the MF solution is still rising reasonably. The MF and the linearized solution split just before the threshold which shows that the linearized approach can be used close to the threshold. Tolerating an error of 5 \%, $\sigma$ is valid from a value of zero up to $0.99895\cdot \sigma_{\mathrm{th}}$

\subsection{Outra-cavity squeezing}
\begin{figure}[b]
	\includegraphics[scale=0.5]{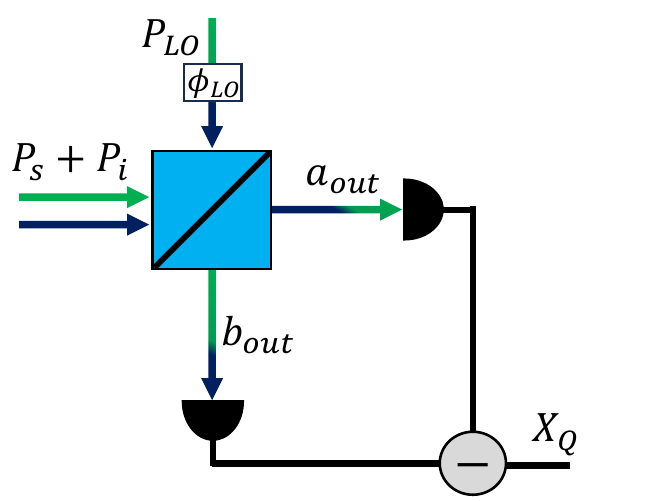}
	\caption{\label{fig:HomodyneMeasurement} Homodyne measurement to realize the measurement of the with the local oscillator consisting of a strong coherent background and a certain phase $\phi_{LO}$ with the same frequencies as $P_s$ and $P_i$.}
\end{figure}
\begin{figure}[b]
	\includegraphics[scale=0.63]{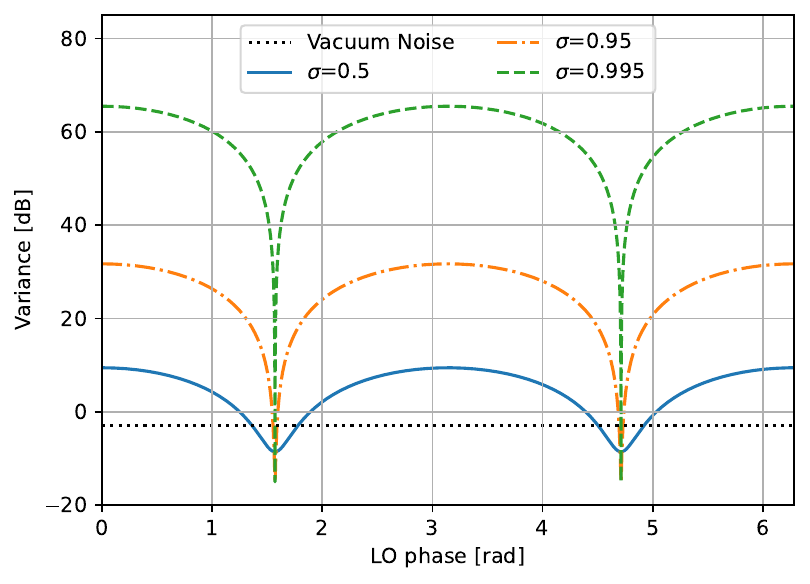}
	\caption{\label{fig:Variance_squeezed} Variance of the two-mode squeezed light calculated using equation \ref{equ:variance} over the LO phase with a variation of $\sigma_n$.}
\end{figure}
\begin{figure}[b]
	\includegraphics[scale=0.63]{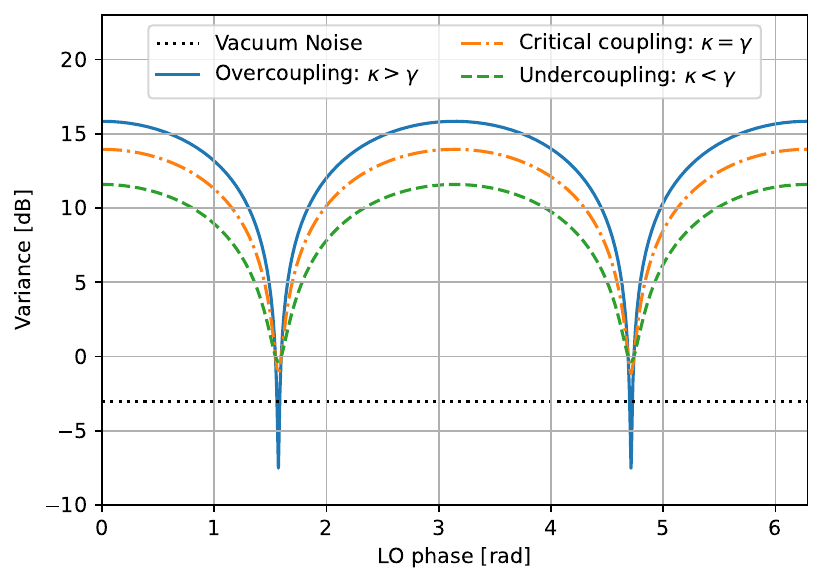}
	\caption{\label{fig:Squeezing_ky} Variance of the two-mode squeezed light calculated using equation \ref{equ:variance} over the LO phase with a variation of $\kappa$ and $\gamma$. The variations are achieved with $\alpha_{loss}=0.23$ 1/m for over-coupling, $\alpha_{loss}=7.28$ 1/m for critical coupling and $\alpha_{loss}=23.04$ 1/m for under-coupling while $\sigma_n$ is constant at 0.95. }
\end{figure}
The squeezing of the outra-cavity modes can be analyzed by using the expectation results determined with the linearized theory of equation \ref{equ:outputMode}. It is of special interest for this work, how the two-mode squeezing of the outra-cavity signal and idler modes behaves depending on the system parameters. One way to analyze the squeezed light is the joint spectral intensity $\Phi$, which depends on the wavelength of the generated signal and idler. It is known from energy conservation of FWM with $2\omega_p=\omega_s+\omega_i$, that if the signal has the detuning $\delta_{\omega}$ from $\omega_p$, the idler has the distance $-\delta_{\omega}$. Consequently, this leads to the definitions of the FWM detuning $\Delta_s=\Delta\omega_s$ as well as $\Delta_i=-\Delta\omega_i$. Thereby, $\Delta \omega_s$ is the distance from $\omega_s$ and $\Delta \omega_i$ the distance from $\omega_i$. The joint spectral intensity between the signal and idler results then in the following equation
\begin{eqnarray}\label{equ:jsi}
\begin{aligned}
\Phi\left(\omega, \omega'\right) &= \langle b_{out,s}^\dagger\left(\omega\right) b_{out,i}^{\dagger}\left(\omega'\right) b_{out,s}\left(\omega\right) b_{out,i}\left(\omega'\right) \rangle  &\\
&=\frac{16\kappa^2\sigma^4\Gamma^2+4\kappa^2\sigma^2(\Lambda+(\Gamma^2+\sigma^2)^2)}{(\Lambda+(\Gamma^2-\sigma^2)^2)^2} \delta(\omega - \omega').&
\end{aligned}
\end{eqnarray}
with $\Lambda=16\Delta\omega_s^2\Delta\omega_i^2+8\Delta\omega_s\Delta\omega_i\sigma^2 + 4\Gamma^2(\Delta\omega_s^2+\Delta\omega_i^2)$.
It can be interpreted as the joint signature probability distribution that indicates the probability to detect a signal photon at a certain frequency if an idler is detected at another frequency. Thus, it is an indication for the correlation between the two modes as a function of the frequency \cite{Zielnicki2018}. The result of $\Phi$ is shown in Figure \ref{fig:ring_squeezing} over the signal and idler frequency with the same values for the geometry as above and a $\sigma_n=0.995$. It is interesting that the spectrum is squeezed with a high correlation and broadband in the case of an asymmetric detuning between the signal and the idler. This behavior can be explained by the energy conservation, which must be fulfilled and means that if the signal is generated at a higher frequency compared to the pump frequency, the idler must be generated at a lower frequency.\\
Another important feature of squeezed light is the quadrature operator which allows the analysis of the noise behavior with \cite{Walls2008}
\begin{equation} \label{equ:quadrature_signal}
	X_Q \left(\omega\right)= \frac{1}{\sqrt{2}}\left[(b_{out,s}\left(\omega\right) + b_{out,i}\left(\omega\right) ) e^{i \phi_{LO}} + \mathrm{h.c.} \right]
\end{equation}
with the local oscillator (LO) phase $\phi_{LO}$. In experiments this phase can be used to change the measured signal from low-noise or squeezing at $\phi_{LO}=\pi/2$ to the high noise or anti-squeezing $\phi_{LO}=0$. Thereby, the outra-cavity light is mixed with a coherent local oscillator laser with a certain phase to realize the desired quadrature operator as shown in figure \ref{fig:HomodyneMeasurement} and experimentally demonstrated in \cite{doi:10.1126/sciadv.aba9186}. This setup is also discussed in detail in appendix \ref{cha:quadMeasurement}. Of special interest is the variance of the quadrature to identify the noise of the two-mode squeezed light which is defined as \cite{Walls2008}
\begin{equation}\label{equ:variance}
\begin{aligned}
	V\left(\omega,\omega'\right) &= \Delta X_Q\left(\omega,\omega'\right) \\ &= \langle X_Q\left(\omega\right) X_Q\left(\omega'\right) \rangle - \langle X_Q\left(\omega\right) \rangle\langle X_Q\left(\omega'\right) \rangle \\
	& \xrightarrow{b_{in,s/i}=b_v} \langle X_Q\left(\omega\right) X_Q\left(\omega'\right) \rangle.
\end{aligned}
\end{equation}
For the case of zero detuning, the following equations are determined for the variance in the squeezed and anti-squeezed case with integrating over $\omega$ to remove the $\delta$ function which leads to
\begin{eqnarray}
	V_{\phi_{LO}=\pi/2}&=& \frac{\Gamma^2+\sigma^2}{(\Gamma-\sigma)^2} - \frac{2\sigma(\kappa-\gamma)}{(\Gamma-\sigma)^2}, \\
	V_{\phi_{LO}=0} &=&  \frac{\Gamma^2+\sigma^2}{(\Gamma+\sigma)^2} +\frac{2\sigma(\kappa-\gamma)}{(\Gamma+\sigma)^2} .
\end{eqnarray}
The results of equation \ref{equ:variance} over the phase $\phi_{LO}$ are shown in Figure \ref{fig:Variance_squeezed} with the same geometry values as above. The different colors and line styles represent various values of the pump power $\sigma_n$ and it can be seen that with rising power, the squeezing improves while the anti-squeezing increases. A squeezing of approximately $-15.02$ dB can be achieved for the defined limit $\sigma_n=0.99895$. This is close to the value with less pump $\sigma_n=0.95$ with $-15.01$ dB which is about $-12$ dB lower than vacuum and thus, good noise reduction can be already achieved with a power close to the threshold. However, the difference for the anti-squeezing is significantly larger with $65.46$ dB for $\sigma_n=0.99895$ and $31.68$ dB for $\sigma_n=0.95$. \\
For the design it is important to know how the ring geometries affect the squeezing. In this case, the two main parameters are the decay rates $\kappa$ and $\gamma$ which can be tuned by the ring design. The influence for different ratios between $\kappa$ and $\gamma$ are shown in Figure \ref{fig:Squeezing_ky}. The best squeezing can be achieved for the case $\kappa>\gamma$ which is also known as over-coupling, since more light couples in and out of the ring than is lost due to losses. For the so called critical coupling case in which both coupling rates are equal with $\kappa=\gamma$, the squeezing decreases and for the under-coupling case with $\kappa<\gamma$ the variance is larger compared to the vacuum noise. It is reasonable that $\kappa>\gamma$ is required for a good squeezing, since only if the squeezed signal and idler pairs couple out of the resonator, they contribute to a squeezing. This matches the results of other works and shows the importance of a proper cavity design \cite{PhysRevA.93.033820}. \\
Another feature that is often given in the literature to describe squeezed light, is the squeezing parameter $r$. By comparison with results for two-mode squeezed light in the literature \cite{Gerry_Knight_2004}, we can derive it to
\begin{equation}
\begin{aligned}
	r& = \sinh^{-1}\left(\sqrt{\langle b_{out,s}^\dagger b_{out,s} \rangle }\right)& \\
	&= \sinh^{-1}\left(\sqrt{\frac{4\sigma^2\kappa\Gamma}{\Xi -2\sigma^2\Gamma^2} }\right).&
\end{aligned}
\end{equation}
Using the same values as for Figure \ref{fig:Variance_squeezed} with $\sigma_n=0.99895$, a squeezing parameter of $r=7.54$ can be achieved.

\section{Mach-Zehnder interferometer}
\label{cha:mach-zehnder}
The main goal of this work is to show the potential of two-mode squeezed states generated by FWM in a sensor application that is built in the form of an MZI and to determine the sensing limit. The considered system from Figure \ref{fig:MZI} is split into a part for the state preparation, the sensor region and the detection. \\
At the state preparation, pump light $P_l$ is send into the ring resonator for the signal and idler generation, while additional coherent light $P_c$ is coupled in the other input of the MZI. The generated outra-cavity signal and idler mode correspond to the state $a_1$ and $P_c$ to $a_0$. They are sent into the MZI via the BS. The states then enter the sensor area, which applies a symmetrical phase shift (PS) to them that is to be measured. The BS and PS can be defined as follows
\begin{eqnarray} \label{bs_ps_matrix}
	\mathbf{BS}  = \frac{1}{\sqrt{2}} \left(\begin{matrix} 
	1 & 1  \\
	1 & -1 
	\end{matrix} \right), 
	\mathbf{PS} = \left(\begin{matrix} 
	e^{i \frac{\phi}{2}} & 0 \\
	0 &e^{-i \frac{\phi}{2}} 
	\end{matrix} \right).
\end{eqnarray}
The sensor region is typically the largest area of the system and thus, it is important to consider losses to model a realistic application. Therefore, the efficiency $\eta$ is introduced, which is related to the mode power and is one for the lossless case. In chip-integrated applications $\eta$ depends on the waveguide losses $\alpha_{loss}$ and the length of the sensor region $L$ with 
\begin{equation}
	\eta = e^{-\alpha_{loss}L}.
	\label{equ:eta_f}
\end{equation}
The derivation of equation \ref{equ:eta_f} is discussed in more detail in appendix \ref{cha:physRel}. To consider the effect of losses, we use a BS that couples with $\eta$ a vacuum mode $b_v$ into the system while coupling out the system mode \cite{DEMKOWICZDOBRZANSKI2015345}. This is discussed in appendix \ref{cha:lossMatrix}. The loss must be applied to each path of the MZI with two uncorrelated vacuum modes $\mathbf{B}=(b_{v,0} \: b_{v,1})^\top$. 
After the sensor region, another BS follows and the modes at the detection stage $\mathbf{D}=(d_0 \: d_1)^\top$ can be determined in dependency of the input modes $\mathbf{A}=(a_0 \: a_1)^\top$ with 
\begin{eqnarray}
	\mathbf{D} = \mathbf{BS} \cdot \left[\sqrt{\eta} \cdot \mathbf{PS} \cdot \mathbf{BS} \cdot \mathbf{A} +
	\sqrt{1-\eta} \cdot \mathbf{B} \right].
\end{eqnarray}
Afterwards, the output modes are detected and further processed with some electronic circuits. For the detection scheme, we use the commonly used intensity difference detection which is defined as 
\begin{equation}
\mathbf{\mathrm{ID}} = d_0^\dagger d_0 - d_1^\dagger d_1.
\end{equation}
It is obvious that in reality losses appear at each stage like at the detector and the electronic processing, which should be modeled using a loss BS \cite{Bachor2004}. As shown in \cite{DEMKOWICZDOBRZANSKI2015345} it is sufficient to focus on the losses within the optical system because these are the most significant ones and the other losses can be merged to this loss.

\section{Quantum-sensing limit}
\label{cha:sensingLimit}

\begin{figure}[b]
	\includegraphics[scale=0.65]{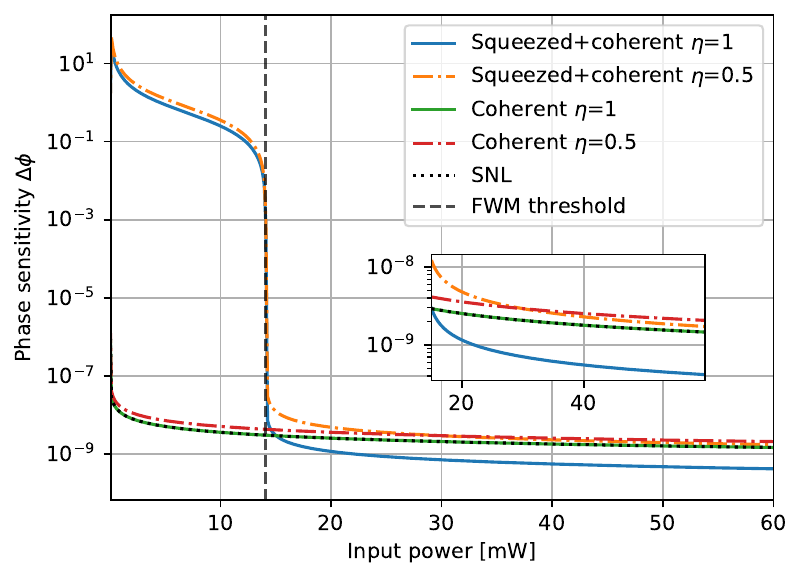}
	\caption{\label{fig:PhaseSensitivity_Comp} Phase sensitivity over the input power for squeezed light mixed with coherent light (blue (solid dark) lossless and orange (dashed dotted light gray) with losses), coherent light (green (solid behind black dots) lossless and red (dashed dotted dark grey) with losses) and the SNL (black dotted). The FWM threshold is shown in black dashed. For the calculation, the following values are used: $\Delta_i=\Delta_s=\Delta_p=0$, $X=0.01$, $n_{\mathrm{eff}}=1.801$, $n_g=2.10087$, $\alpha_{loss}=0.23$ $1/\mathrm{m}$, $L=2\pi\cdot220\cdot10^{-6}$ m, $\lambda_p=1550\cdot10^{-6}$ m, $n_2=2.4\cdot10^{-19}$  $\mathrm{m}^2$/W, $A_{\mathrm{eff}}=1.05564\cdot10^{-12}$ $\mathrm{m}^2$.}
\end{figure}

To derive the performance and the quantum-enhancement of the system, it is necessary to determine the noise of the measurement operator. In general the minimum detectable phase change inside of the MZI $\Delta \phi$ can be evaluated using error propagation \cite{PhysRevLett.72.3439, DEMKOWICZDOBRZANSKI2015345}
\begin{eqnarray}
	\Delta \phi = \frac{\sqrt{\Delta \mathrm{ID} }}{|\delta \langle \mathrm{ID}  \rangle / \delta \phi|}
	\label{equ:phase_limit}
\end{eqnarray}
with $\Delta \mathrm{ID} = \langle \mathrm{ID}^2 \rangle - \langle \mathrm{ID} \rangle^2$. Therefore, the previous derived expectation values are required. Since higher order expectation values appear in equation \ref{equ:phase_limit}, we use the helpful cumulant expansion which decreases the order of an operator $\mathbf{X}=\langle X_1 X_2 ... X_n \rangle$ \cite{Kubo1962}
\begin{equation}
	\langle X_1 X_2 ... X_n \rangle = \sum_{p \in P(I)/I} (|p|-1)! (-1)^{|p|} \prod_{B \in p} \langle \prod_{i \in B} X_i\rangle
\end{equation}
with $I=\left\{1, 2, ..., n \right\}$, $P(I)/I$ being the set of all operators excluding $I$, $|p|$ the length of the partition $p$ and $B$ is the block of each partition. An example for $n=3$ is
\begin{equation}
\begin{aligned}
	\langle X_1 X_2 X_3 \rangle &= \langle X_1 X_2 \rangle \langle X_3 \rangle + \langle X_1 X_3 \rangle \langle X_2 \rangle \\
	&+ \langle X_1 \rangle \langle X_2 X_3 \rangle - 2 \langle X_1 \rangle \langle X_2 \rangle \langle X_3 \rangle.
\end{aligned}
\end{equation}
Using this, it is possible to determine the noise of the measured signal and thus, to find out the minimal detectable phase change of the system. \\
In appendix \ref{cha:mzi_phaseSensitivity}, it is shown that the optimal phase sensitivity in our setting can be achieved for a phase difference of $\phi_{opt}=\pi/2$ between the two paths of the MZI. Thus, the results in the following are evaluated at this operation point. If the MZI is operated using only $P_c$ with the mode amplitude $\alpha_c$ and without using $P_l$ for the squeezd light generation, the phase sensitivity using only coherent light in the sensor region $\Delta\phi_{\mathrm{c}}$ is given by the following expression
\begin{equation}
\Delta\phi_{\mathrm{c}} = \frac{1}{\sqrt{\eta}\alpha_c}.
\label{equ:phaseSense_c}
\end{equation}
Equation \ref{equ:phaseSense_c} scales as good as the shot-noise limit (SNL) with $1/\alpha_c$ for the lossless case and with losses it is slightly worse. For the case of using $P_l$ to pump the ring resonator to generate squeezed light and mixing this with coherent light $P_c$, the phase sensitivity using squeezed light $\Delta\phi_{\mathrm{s}}$ at $\phi_{opt}$ can be expressed as
\begin{widetext}
	\begin{equation}
	\Delta\phi_{\mathrm{s}} = \frac{\sqrt{\eta\alpha_c^2(\Gamma-\sigma)^2(\Gamma^2+\sigma(2\gamma-6\kappa)+\sigma^2)+\alpha_c^2(\Gamma^2-\sigma^2)^2+8\kappa\sigma^2\Gamma}}{\sqrt{\eta}(\Gamma^2-\sigma^2) |\alpha_c^2-\frac{8\sigma^2\kappa\Gamma}{(\Gamma^2-\sigma^2)^2}|}.
	\label{equ:phase_sensitivity_squeezed}
	\end{equation}
\end{widetext}
Two poles can appear. One at $\sigma=\Gamma$, which is however not in the valid range of $\sigma$ and the second one if the number of squeezed photons $\langle b_{out,s}^\dagger b_{out,s} \rangle + \langle b_{out,i}^\dagger b_{out,i} \rangle$, given by equation \ref{equ:photon_n}, matches the number of coherent photons $\alpha_c^2$. This can be seen in the denominator of equation \ref{equ:phase_sensitivity_squeezed}. However, this only appears for rather small values of $\alpha_c$, since the number of squeezed photons is limited in practice. Thus, $\alpha_c$ should be chosen large enough, which in most applications, is easy reachable since the number of squeezed photons is small. This behavior is analyzed in more detail in appendix \ref{cha:mzi_phaseSensitivity}.\\
For the optimal conditions, with no losses and thus with $\eta=1$, $\gamma=0$ and a large amount of coherent photons with $\alpha_c > 10^4$, the phase sensitivity of equation \ref{equ:phase_sensitivity_squeezed} can be approximated by the following expression
\begin{equation}
\Delta\phi_{\mathrm{s, opt}} \approx \frac{\sqrt{2}}{ \alpha_c} \frac{\kappa-\sigma}{\kappa+\sigma}.
\label{equ:phaseSense_sOpt}
\end{equation}
Depending on the degree of squeezing described by the values for $\kappa$ and $\sigma$, an improvement in scaling over equation \ref{equ:phaseSense_c} and the SNL can be achieved by up to three orders of magnitude within the valid range of $\sigma$. 
However, equation \ref{equ:phaseSense_sOpt} is challenging to achieve in a real world application, since losses will appear. \\
To identify how well both systems perform in a more realistic and lossy setting, it is necessary to also analyze the SNL of the system. This SNL is calculated with the total number of photons inside of the MZI. However, as indicated in the state preparation stage in Figure \ref{fig:MZI}, the pump power that is required to generate the two-mode squeezed light is included as well in the SNL. This results in the following equation for the SNL
\begin{equation}
\Delta\phi_{\mathrm{SNL}} = \frac{1}{\sqrt{N}} =\frac{1}{\sqrt{\langle d_0^\dagger d_0 \rangle+\langle d_1^\dagger d_1 \rangle  + \left|\alpha_l \right|^2}}.
\label{equ:snl}
\end{equation}
with the number of photons in the system $N$. The addition of $\alpha_l$ is required since the pump is not modeled in the outra-cavity expectation values. In our opinion, it is important to include $\alpha_l$ in equation \ref{equ:snl}. The reason is that $\alpha_l$ is used inside the sensor system to pump the ring resonator and to generate the squeezed light and thus, it already takes part in the whole sensor performance. Pumping the ring resonator only makes sense if the generation of squeezed light is more advantageous than using the pump power directly for the phase detection in the sensor region. This fact can be made more comparable if $\alpha_l$ is included in the equation \ref{equ:snl}. \\
To analyze this behavior in more detail, the phase sensitivity is shown as a function of the input power in Figure \ref{fig:PhaseSensitivity_Comp} with coherent input of equation \ref{equ:phaseSense_c}, squeezed light mixed with coherent light of equation \ref{equ:phase_sensitivity_squeezed} and the SNL of equation \ref{equ:snl} for comparison. As expected, it can be seen that the phase sensitivity for a coherent input improves with increasing power by the same scaling as the SNL. Even with a higher loss and thus, a lower $\eta$, the scaling is similar and just the performance decreases slightly by a factor of $1 / \sqrt{\eta}$. \\
In case of using squeezed light, $P_l$ is increased close to threshold with $\sigma_n=0.99895$ which corresponds to a power of $P_l=14.12$ mW, while $P_c=0$. Note that $P_l$ does not contribute to the sensor performance but to the SNL as indicated by equation \ref{equ:snl}. Instead, $P_l$ is required to pump the ring resonator and generates only a weak two-mode squeezed light that is available for sensing. The generated signal and idler power at $\sigma_n=0.99895$ is $P_s=P_i\approx1.13\cdot10^{-10}$ mW, which is very low compared to the input power. With increasing power and thus a larger $\sigma_n$, the sensitivity $\Delta\phi_{\mathrm{s}}$ increases. However, the performance by using coherent light is still much better than using squeezed light, which can achieve a phase sensitivity of about $0.008$. By keeping $P_l$ constant at 14.12 mW close to the threshold and increasing $P_c$, the performance is improved significantly and the phase sensitivity of the squeezed light mixed with coherent light surpasses the coherent as well as the SNL performance. This is possible just due to the small amount of squeezed light.\\
However, the impact of losses on squeezed light is much larger than on coherent light, which is in agreement with \cite{SCHNABEL20171}. Despite this fact, the scaling over input power is still better for the lossy case of the squeezed light mixed with coherent light and thus, it surpasses the lossy coherent performance and would also improve further compared to the SNL. This is shown in the zoomed part of Figure \ref{fig:PhaseSensitivity_Comp}. \\
In order to achieve useful quantum enhancement and an advantage over the exclusive use of classical light, two essential requirements must be met. The first one is that a low threshold power needs to be achieved, because the scaling of the squeezed light mixed with coherent light is superior to using only coherent light. However, this advantage only applies if the threshold allows the squeezed light sensitivity to surpass the coherent sensitivity at a reasonable power. The threshold can be changed by adapting the design, as shown in equation \ref{equ:photon_n}, or by using a material with a high-nonlinear susceptibility.\\
\begin{figure}[b]
	\includegraphics[scale=0.68]{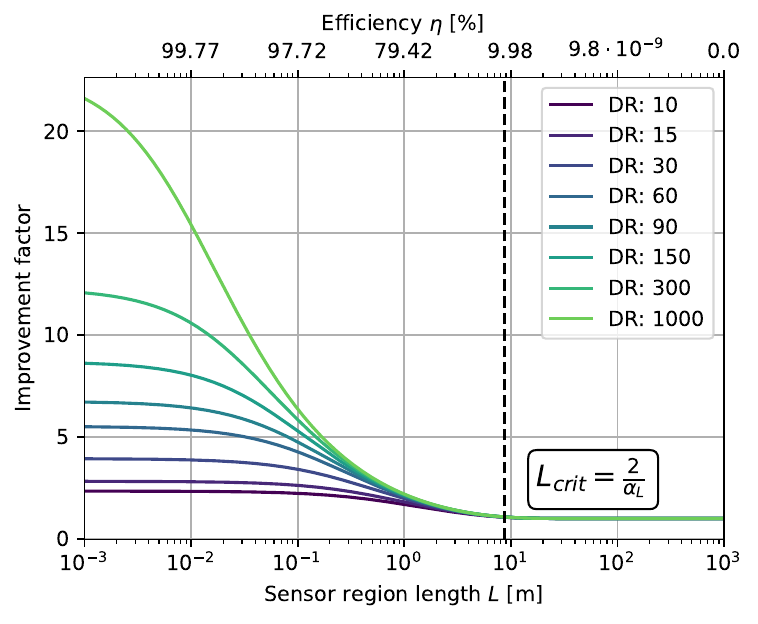}
	\caption{\label{fig:Sagnac_Performance} Improvement factor evaluated with equation \ref{equ:improvement} over the waveguide length of the sensor region for different decay ratios from dark to light for a rising decay ratio. For the calculation, the following values are used: $\Delta_i=\Delta_s=\Delta_p=0$, $X=0.01$, $n_{\mathrm{eff}}=1.801$, $n_g=2.10087$, $\alpha_{loss}=0.23$ $1/\mathrm{m}$, $L=2\pi\cdot220\cdot10^{-6}$ m, $\lambda_p=1550\cdot10^{-6}$ m, $n_2=2.4\cdot10^{-19}$  $\mathrm{m}^2$/W, $A_{\mathrm{eff}}=1.05564\cdot10^{-12}$ $\mathrm{m}^2$.}
\end{figure}
The second requirement is that the losses inside of the sensor region needs to be below a critical value. In chip-integrated optical sensors, these losses depend mostly on the scale of the sensor region with equation \ref{equ:relPowChange}. A smaller sensor region may correspond to lower losses, but also a weaker signal since the light requires a certain interaction length with the stimulus. Examples for this are Sagnac interferometers in which a larger sensing area leads to a stronger phase shift or Raman measurements, in which the interaction length between the light and the molecules enhances the signal. To analyze the performance improvement of squeezed light compared to coherent light, we define the improvement factor $\mathbbm{I}$ which is the ratio between $\Delta \phi_c$ and $\Delta \phi_s$ with
\begin{equation}
	\mathbbm{I} = \frac{\Delta \phi_c}{\Delta \phi_s}, 
	\label{equ:improvement}
\end{equation}
Since the squeezing depends on the decay rates $\kappa$ and $\gamma$ as it is shown in Figure \ref{fig:Squeezing_ky}, equation \ref{equ:improvement} is analyzed for a variation of the ratio of both which we define as the decay ratio with 
\begin{equation}
\mathrm{DR}=\frac{\kappa}{\gamma}.
\end{equation}
The results of equation \ref{equ:improvement} are shown in Figure \ref{fig:Sagnac_Performance} over the length of the sensor region $L$ as well as the efficiency $\eta$. It can be seen that for short lengths and rising decay ratios, the improvement factor rises as well which proves the quantum advantage. The improvement factor for a short length and a high $\eta$ can be approximated with
\begin{equation}
\mathbbm{I}_{\mathrm{Length}\rightarrow 0}=10.218\cdot \mathrm{ln}\biggl( \mathrm{DR}+143.47 \biggr)-49.816.
\end{equation}
In contrast to this is the behavior with long lengths where $\mathbbm{I}$ reaches one. This can be explained by the fact that losses increase with length and the squeezed state is highly susceptible to them. After a certain length most of the squeezing is lost and no advantage can be achieved. We call this length the critical length which depends on the losses and can be approximated as 
\begin{equation}
L_{crit} = \frac{2}{\alpha_{loss}}
\end{equation}
which corresponds to an efficiency $\eta\approx13.5\%$. For a useful improvement factor of more than 1.5, the efficiency needs to be above $60 \%$. This behavior confirms our starting motivation, that the quantum-enhancement shows its strength for chip-integrated applications which mostly consist of short lengths and low-loss applications. Dependent on the use case, it can be useful to decrease the size of a sensor and to increase the sensitivity using two-mode squeezed light.

\section{Conclusion}
In this work we analyzed the generation of two-mode squeezed light in a ring-resonator cavity and its application for sensing a phase shift in an MZI. The results show that the combination of squeezed and coherent light in a low-loss sensor region can improve the sensitivity compared to the SNL and a classical sensing system. We started with a linearized Hamiltonian for the FWM process and derived the expectation values of the outra-cavity signal and idler modes via the input-output theory. They are used to determine the squeezing behavior of the two-mode squeezed light and to give a support to design a cavity for the desired squeezing behavior. Additionally, the derived equations are used to calculate the minimum detectable phase change inside of an MZI and it has been shown that squeezed light mixed with coherent light can surpass the phase sensitivity of the SNL. Thus, equations are derived which describe the performance of the whole system consisting of a ring resonator for the squeezed light generation in addition with the MZI and ID detection. \\
However, it is important for a real use case to tune the threshold by design and material choice in a reasonable region, where the improved slope of the squeezed light can show its full potential. Furthermore, we showed that the losses of a system limit the usability of two-mode squeezed light. The reason is that squeezed light is very loss sensitive and after a certain length of the sensor region, most of the squeezing is lost and no improvement can be achieved compared to a classical sensor. However, if the sensor region is small and has low losses, two-mode squeezed light can significantly increase the performance by more than a factor of ten. It was shown that this improvement factor depends on the decay ratio and thus, on the design of the cavity. This leads to an optimization problem between a desired decay ratio and a low threshold. If both are reasonably chosen, the quantum enhancement can be beneficial.\\
In summary we think that this can lead to more compact sized optical sensors which can have the same sensitivity at a small size as with a larger sensor region due to the quantum enhancement. It is important to note that squeezed light can improve the performance of many various optical sensor concepts and, as shown in this work, especially for interferometer based sensor principles that are nowadays already available to manufacture and to use.

\appendix

\section{Gain of the four-wave mixing process}
\label{cha:fwm_gain}

Based on the derivation in \cite{PhysRevA.92.033840} the gain of the FWM-process inside of a ring resonator can be given in the following expression
\begin{equation}
	g = \frac{\hbar (\omega_p^2 \omega_s \omega_i)^{1/4} v_g^2 \gamma_{NL}}{L}
\end{equation}
with the group velocity $v_g$, the length of the resonator $L$ and the nonlinear waveguide parameter $\gamma_{NL}$ which is given by \cite{AGRAWAL201327}
\begin{equation}
	\gamma_{NL} = \frac{\omega_p n_2}{c A_{\mathrm{eff}}}
\end{equation}
with the effective mode area of the optical waveguide mode $A_{\mathrm{eff}}$ and the nonlinear refractive index $n_2$ which can be described as 
\begin{equation}
	n_2 = \frac{3\chi^{(3)}}{4 \varepsilon_0 c n_{\mathrm{eff}}^2}
\end{equation}
and the effective refractive index $n_{\mathrm{eff}}$. Since $n_2$ is mostly used in the literature we give our final gain as following
\begin{equation}
	g = \frac{\hbar \omega_p (\omega_p^2 \omega_s \omega_i)^{1/4} v_g^2 n_2}{c A_{\mathrm{eff}}L} \approx \frac{\hbar \omega_p^2 v_g^2 n_2}{c A_{\mathrm{eff}}L}.
\end{equation}
Note that the derivations slightly differ from each other as in \cite{Hoff:15}. Since it is derived with some approximations, the gain can only be confirmed by experiments.

\section{Physical relation to the parameters}
\label{cha:physRel}
For a practical realization of a system it is very important to know the physical relation of the operators and how they depend on the structure dimensions. The parameters for the design of the ring resonator are the decay rate $\gamma$ and the coupling rate $\kappa$. They describe how much light decays via the respective effect at each round trip time which is described as follows for a laser pulse
\begin{equation}
	t_{round} = \frac{n_{\mathrm{eff}} L}{c}
\end{equation}
with the effective refractive index of the waveguide $n_{\mathrm{eff}}$, the length of the ring $L$ and the speed of light $c$. At each round trip a certain part of the light couples out of the ring resonator into the connected waveguide by a certain fraction and this is described by the cross coupling $X$ which takes the values $0 \leq X \leq 1$. All the light is coupled to the waveguide at a value of one and no light at a value of zero. Thus, the coupling rate can be defined as the light that is coupled out per round trip
\begin{equation}
	\kappa = \frac{X}{t_{round}}  =  \frac{X c}{n_{\mathrm{eff}} L}.
\end{equation}
Additionally, at each round-trip a certain part of the light is scattered away caused by for example scattering on rough waveguide walls. In general, the optical power $P$ in lossy photonic chip-integrated waveguides over a length $L$ can be described by the following equation \cite{Hunsperger2002}
\begin{equation}
P(L) = P_0 \cdot e^{-\alpha_{loss} L}
\end{equation}
with the optical waveguide losses $\alpha_{loss}$ in the units of $m^{-1}$. To model the losses now in dependency of the waveguide losses and together with the knowledge that $P(L) \leq P_0$, we simply define it as the relative power change with a value of one for no losses and a value of zero for maximum losses
\begin{equation}
\eta =1-\frac{P_0 - P(L)}{P_0} = e^{-\alpha_{loss} L}. 
\label{equ:relPowChange}
\end{equation}
Based on this we can define the decay rate per round trip time as follows
\begin{equation}
	\gamma = \frac{1-\eta}{ t_{round}} = \frac{(1 - e^{-\alpha_{loss} L}) c}{n_{\mathrm{eff}} L}.
\end{equation}
Another important physical parameter is the complex amplitude of a coherent pump. It is known that this can be described as 
\begin{equation}
	\alpha_l = \sqrt{\frac{P}{\hbar \omega_p}}\cdot \exp(i\phi_{l})
	\label{equ:photon_amplitude}
\end{equation}
with the pump power $P$ and the frequency of the light $\omega_p$ and is proportional to a photon flux. Note that $\alpha_l$ has the unit of $\sqrt{\mathrm{Hz}}$. However, the intra-cavity amplitude of equation \ref{equ:cavity_pumpPhotons} $\alpha_p$ in the main text is unit-less and cannot be converted back to the optical power using equation \ref{equ:photon_amplitude}. To achieve this, it is necessary to transfer the units to Hertz which is done with defining the transmission rate with
\begin{equation}
	t = \frac{(1-X) c}{n_{\mathrm{eff}} L}.
\end{equation}
This is just the counterpart of the coupling rate and defines the rate of photons staying either in the resonator or in the straight waveguide. The intra-cavity amplitude in units of Hertz can then be derived as $a_{p,\sqrt{\mathrm{Hz}}} = \sqrt{t} \cdot a_{p}$. \\
Another important parameter for the FWM process is the injection parameter $\sigma$ and its relation to the FWM threshold. The physical dependency for the injection parameter can be given as
\begin{equation}
	\sigma = \frac{2g\kappa}{\left( \Gamma/2 - i\Delta\right)^2}\frac{P_l}{\hbar \omega_p}.
\end{equation}
It is possible to derive the FWM threshold power that is required in the waveguide $P_l$ in Figure \ref{fig:ring_squeezing} by using the relation $\sigma_{\mathrm{th}}=\Gamma$ and equation \ref{equ:photon_amplitude}. This leads to
\begin{equation}\label{equ:thresholdPower}
	P_{\mathrm{th}} = \frac{\Gamma\hbar\omega_p\left( \Gamma/2 - i\Delta_p\right)^2 }{2g\kappa} \overset{\Delta=0}{=} \frac{\Gamma^3 \hbar\omega_p}{8g_\mathrm{opt}\kappa}.
\end{equation}
Note, that effects like self-phase and cross-phase modulation are neglected in equation \ref{equ:thresholdPower} and thus, the real $P_{\mathrm{th}}$ might be higher. 

\section{Quadrature measurement}
\label{cha:quadMeasurement}

In the following, the determination of the quadrature measurement shown in figure \ref{fig:HomodyneMeasurement} is explained in more detail. One input of the BS consists of the generated signal and idler $a_{BS,1} = a_s\cdot \exp\left(-i\left[\omega_{s}t+\phi_s\right]\right) + a_i\cdot \exp\left(-i\left[\omega_{i}t+\phi_i\right]\right) $ with a certain phase $\phi_s, \phi_i$ for the signal and idler. The other input consists of a local oscillator with a strong coherent background that includes waves of the signal and idler frequency with $a_{BS,2} = |\alpha_{LO}| \cdot \left( \exp\left(-i\left[\omega_{s}t+\phi_{LO}\right]\right) + \exp\left(-i\left[\omega_{i}t+\phi_{LO}\right]\right) \right)$ and the phase $\phi_{LO}$. In the following, we assume that the signal and idler phase matches with $\phi_i=\phi_s$. Both inputs are mixed at the beam splitter with the matrix from equation \ref{bs_ps_matrix}, detected and the measured signal is subtracted which leads to
\begin{equation}
\begin{aligned} \label{homodyne_detection}
	X_Q &= d_1^\dagger d_1 - d_2^\dagger d_2\\
	 &= |\alpha_{LO}| \cdot \left( a_s e^{-i\left(\delta_{LO} + t\left[\omega_{s}-\omega_{i}\right]\right)} + a_s^\dagger e^{i\left(\delta_{LO} + t\left[\omega_{s}-\omega_{i}\right]\right)} \right) \\
	 &+|\alpha_{LO}| \cdot \left(  a_i e^{i\left(-\delta_{LO} + t\left[\omega_{s}-\omega_{i}\right]\right)} + a_i^\dagger e^{-i\left(-\delta_{LO} + t\left[\omega_{s}-\omega_{i}\right]\right)} \right)	 \\
	 &+ |\alpha_{LO}| \left( \left[a_i + a_s \right]\cdot e^{-i\delta_{LO}} + \left[a_i^\dagger + a_s^\dagger \right] e^{i\delta_{LO}} \right)
\end{aligned}	
\end{equation}
with the difference between the signal, idler and the LO phase $\delta_{LO} = \phi_i-\phi_{LO}$. The equation \ref{homodyne_detection} includes one fast oscillating and one static term. Since the fast oscillating one cannot be measured by a physical detector, the static one is measured which corresponds to the quadrature signal of equation \ref{equ:quadrature_signal}. Note, that the local oscillator is required in this measurement scheme. The reason is that squeezed light consists generally only of a low amount of photons. Using a local oscillator with a strong coherent background, amplifies the quadrature information of the squeezed light to make it measurable.

\section{Outra-cavity expectation values}
\label{cha:expResults}

In the following, the results of the calculated outra-cavity expectation values are given and determined based on equation \ref{equ:outputMode}. 
To keep the results general, we determine the expectation values for the case that the signal and idler input is in the vacuum state and for the other case that they are in a coherent state with $b_{in,s}=\alpha_{s}$ and $b_{in,i}=\alpha_{i}$. Therefore, the expectation values are split into a static $\langle b_{st} \rangle$ and a fluctuating part $\langle b_{f} \rangle$ with the total expectation value being $\langle b_{out} \rangle = \langle b_{f} \rangle + \langle b_{st} \rangle$. If a vacuum input is assumed then only the fluctuation results are of interest as it is done in the main text. Otherwise, for a coherent input, $\langle b_{f} \rangle$ and $\langle b_{st} \rangle$ needs to be added. However, only the signal modes and some cross products between the signal and idler modes are given. The reason is that due to the symmetry in FWM, the idler results can be simply achieved by a replacement of the index in the equations. Based on them, all desired expectation values can be determined.

\subsection{Fluctuating results}

For the fluctuating results with $\alpha_{i/s}=0$, all first order expectation values are zero $ \langle b_{s,f} \left(\omega \right) \rangle = \langle b_{s,f}^\dagger \left(\omega \right) \rangle = 0 $ as well as the following second order values: $	\langle b_{s,f}\left(\omega \right) b_{s,f} \left(\omega' \right)\rangle = \langle b_{s,f}^\dagger \left(\omega \right)b_{s,f}^\dagger \left(\omega' \right) \rangle =  \langle b_{i,f} \left(\omega \right)b_{s,f}^\dagger\left(\omega' \right) \rangle =  \langle b_{s,f} \left(\omega \right)b_{i,f}^\dagger \left(\omega' \right)\rangle =  \langle b_{s,f}^\dagger \left(\omega \right) b_{i,f} \left(\omega' \right) \rangle =  \langle b_{i,f}^\dagger \left(\omega \right) b_{s,f} \left(\omega' \right)\rangle = 0 $. The relevant results are given in the following:
\begin{widetext}
	\begin{equation}
	\langle b_{i,f} \left(\omega \right)b_{s,f} \left(\omega' \right)\rangle = \frac{-2\kappa\sigma(4\Delta_i\Delta_s-2i\Gamma(\Delta_i+\Delta_s)-\Gamma^2-\sigma^2)}{\Xi - 2\sigma^2\Gamma^2} \delta(\omega - \omega')
	\end{equation}
	\begin{equation}
	\langle b_{i,f}^\dagger\left(\omega \right) b_{s,f}^\dagger \left(\omega' \right)\rangle = \frac{-2\kappa\sigma(4\Delta_i\Delta_s+2i\Gamma(\Delta_i+\Delta_s)-\Gamma^2-\sigma^2)}{\Xi - 2\sigma^2\Gamma^2} \delta(\omega - \omega')
	\end{equation}
\end{widetext}
The equation for the photon number $\langle b_{s,f}^\dagger\left(\omega \right) b_{s,f}\left(\omega' \right) \rangle$ is already given in equation \ref{equ:photon_n}. Note that the other expectation values can be determined either due to the relation $\langle b_{s,f}\left(\omega \right) b_{s,f}^\dagger \left(\omega' \right)\rangle = 1+\langle b_{s,f}^\dagger\left(\omega \right) b_{s,f} \left(\omega' \right)\rangle$ or due to symmetries like $\langle b_{i,f}^\dagger\left(\omega \right) b_{s,f}^\dagger \left(\omega' \right)\rangle = \langle b_{s,f}^\dagger\left(\omega \right) b_{i,f}^\dagger\left(\omega' \right) \rangle$. Thus, all expectation values for the fluctuating results can be determined with the few given equations.

\subsection{Static results}
For the static results each expectation value is unequal zero. However, to determine each expectation value only the following two equations \ref{equ:static_exp} and \ref{equ:static_exp_d} are required and the rest can be determined using symmetries.
\begin{figure}[b]
	\includegraphics[scale=0.55]{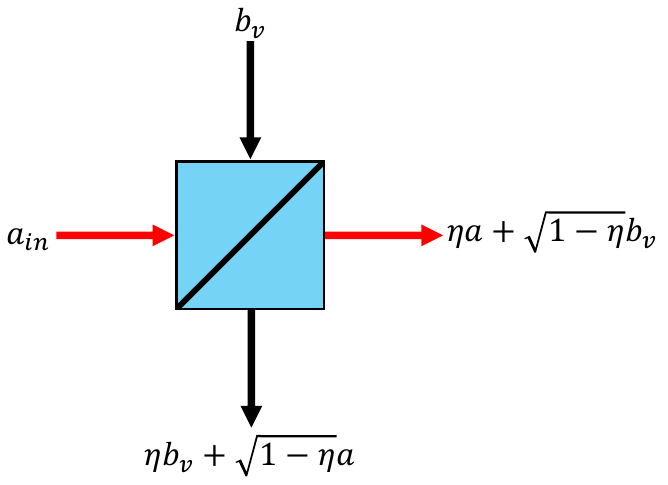}
	\caption{\label{fig:LossBS} Schematic of a BS that models the losses. }
\end{figure}
\begin{figure}[b]
	\includegraphics[scale=0.66]{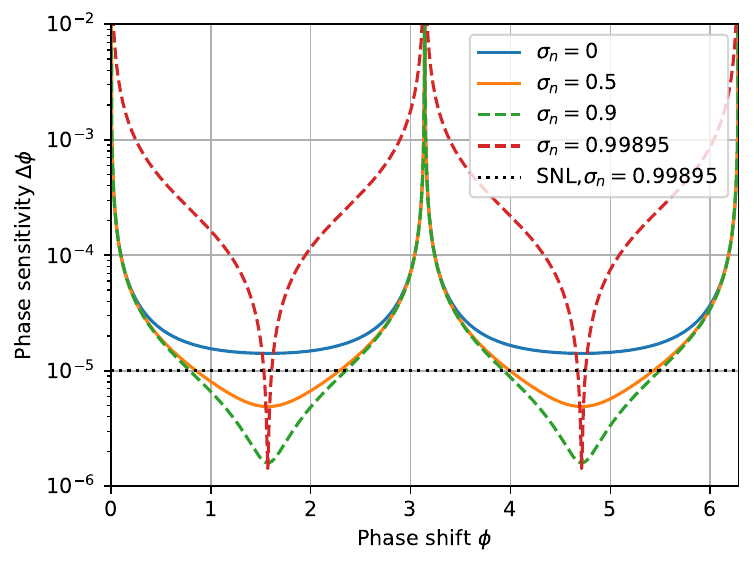}
	\caption{\label{fig:MZI_PhaseSensitivity} Phase sensitivity over the MZI phase shift $\phi$ with a weak coherent pump $\alpha_c=10^5 \sqrt{\mathrm{Hz}}$ and for different $\sigma_n$.}
\end{figure}
\begin{widetext}
	\begin{equation}
	\langle b_{s,st} \left(\omega\right)\rangle =  \frac{\alpha_{s}(\kappa^2+\sigma^2-\gamma^2-4\Delta_i\Delta_s-2i[\Delta_i(\gamma-\kappa)+\Delta_s\Gamma])-2\alpha_{i}\kappa\sigma}{4\Delta_i\Delta_s+2i\Gamma(\Delta_i-\Delta_s)+\Gamma^2-\sigma^2}
	\label{equ:static_exp}
	\end{equation}
	\begin{equation}
	\langle b_{s,st}^\dagger\left(\omega \right) \rangle =  \frac{\alpha_{s}(\kappa^2+\sigma^2-\gamma^2-4\Delta_i\Delta_s+2i[\Delta_i(\gamma-\kappa)+\Delta_s\Gamma])-2\alpha_{i}\kappa\sigma}{4\Delta_i\Delta_s-2i\Gamma(\Delta_i-\Delta_s)+\Gamma^2-\sigma^2}
	\label{equ:static_exp_d}
	\end{equation}
\end{widetext}
The other equations can be simply determined by a factorization with
\begin{figure}[b]
	\includegraphics[scale=0.66]{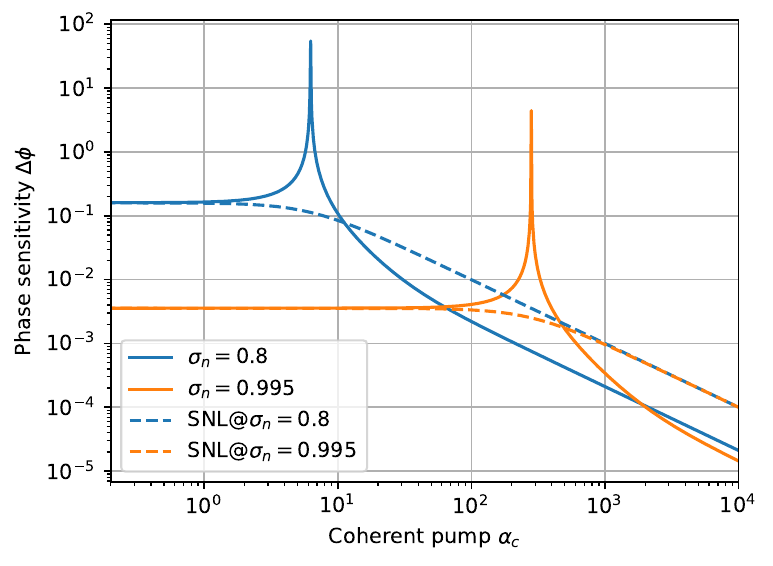}
	\caption{\label{fig:MZI_polePosition} Phase sensitivity over the coherent pump $\alpha_c$ at MZI phase shift $\phi=\pi / 2$ for different $\sigma_n$.}
\end{figure}
\begin{eqnarray}
\langle b_{s,st}\left(\omega \right) b_{s,st}^\dagger \left(\omega' \right)\rangle = \langle b_{s,st} \left(\omega \right)\rangle \cdot \langle b_{s,st}^\dagger \left(\omega' \right)\rangle \\
\langle b_{s,st}^\dagger\left(\omega \right) b_{s,st} \left(\omega' \right)\rangle = \langle b_{s,st}^\dagger \left(\omega \right)\rangle \cdot \langle b_{s,st}\left(\omega' \right) \rangle \\
\langle b_{i,st} \left(\omega \right) b_{s,st} \left(\omega' \right) \rangle = \langle b_{i,st} \left(\omega \right) \rangle \cdot \langle b_{s,st}\left(\omega' \right) \rangle
\end{eqnarray}
or by changing the index between the signal and idler modes. In this way, all static expectation values can be determined.

\section{Waveguide losses in the MZI}
\label{cha:lossMatrix}
Losses in quantum systems are mostly modeled using a BS description with an efficiency $\eta$ as following
\begin{eqnarray}
\mathbf{LM}  = \left(\begin{matrix} 
\sqrt{\eta} & \sqrt{1-\eta}  \\
\sqrt{1-\eta} & -\sqrt{\eta}
\end{matrix} \right)
\label{equ:loss_M_a}
\end{eqnarray}
Thereby, $\eta$ is in the range $0 \leq \eta \leq 1$ with a value of $\eta=1$ for no losses and is described by the equation \ref{equ:relPowChange}. It is important to note that while a part of a waveguide mode $a_{in}$ is lost, vacuum is coupled into the system \cite{Bachor2004}. This is shown in Figure \ref{fig:LossBS} and the resulting mode can be determined by
\begin{equation}
	a_{out} = \sqrt{\eta} a_{in} + \sqrt{1-\eta} b_v
\end{equation}
and the vacuum bath mode $b_v$. The part that couples out of the waveguide can be described by \ref{equ:loss_M_a}. The second output of our virtual BS is ignored.

\section{MZI Phase sensitivity}
\label{cha:mzi_phaseSensitivity}
In the following, the phase sensitivity for the MZI in Figure \ref{fig:MZI} and equation \ref{equ:phase_limit} is discussed in more detail. The behaviour over the phase shift inside of the MZI is analysed first and shown in Figure \ref{fig:MZI_PhaseSensitivity} for a weak coherent signal $\alpha_c=10^5$ and for different pump power $\sigma_n$. It can be seen that the sensitivity diverges for a phase of $\phi=0$ and $\phi=\pi/2$ while it reaches the optimum at $\phi=\frac{\pi}{2}$ and $\phi=\frac{3\pi}{2}$ which should be the preferred working points for a sensing application that can surpass the SNL. Additionally, the sensitivity increases together with $\sigma_n$. This is clear since the squeezing increases as well with $\sigma_n$. In comparison, also the SNL using equation \ref{equ:snl} with $\alpha_p=0 \sqrt{\mathrm{Hz}}$ is shown for the case with the $\sigma_n=0.99895$. As expected, the phase-sensitivity for the same pump surpasses the SNL by more than a magnitude. This shows that two-mode squeezed light can be used for quantum-enhanced metrology which matches other work \cite{OleSteuernagel_2004}.\\
Additionally, it is interesting to analyze the pole-point that results out of equation \ref{equ:phase_sensitivity_squeezed}. Therefore, equation \ref{equ:phase_sensitivity_squeezed} is plotted over the coherent pump $\alpha_c$ for two different $\sigma_n$ values in Figure \ref{fig:MZI_polePosition}. It can be seen that for both $\sigma_n$ values, a pole is visible which is similar to the results of other works \cite{PhysRevA.100.063821}. Interestingly, the position of this pole-point appears if the number of photons of the coherent pump $|\alpha_c|^2$ matches the number of photons of the squeezed states $\langle a_s^\dagger a_s \rangle + \langle a_i^\dagger a_i \rangle$. The phase sensitivity of the MZI is constant up to the pole-position and improves afterwards with rising $\alpha_c$. For both $\sigma_n$ values, the improvement is similar after the pole while the curve using the higher $\sigma_n$ surpasses the lower one. However, before the pole-point, the phase sensitivity is the same as the SNL. This matches the results of \cite{PhysRevLett.100.073601} and shows that a high coherent pump is required to achieve an improved scaling compared to the SNL.

\begin{acknowledgments}
	We are very grateful for the valuable discussions with Carlos Navarrete-Benlloch.\\\\
	Funding: The IPCEI ME/CT project is supported by the Federal Ministry for Economic Affairs and Climate Action on the basis of a decision by the German Parliament, by the Ministry for Economic Affairs, Labor and Tourism of Baden-W\"urttemberg based on a decision of the State Parliament of Baden-W\"urttemberg, the Free State of Saxony on the basis of the budget adopted by the Saxon State Parliament, the Bavarian State Ministry for Economic Affairs, Regional Development and Energy and financed by the European Union - NextGenerationEU.
\end{acknowledgments}

\bibliography{Quant_Bib}

\end{document}